# Innovative Research on IoT Architecture and Robotic Operating Platforms: Applications of Large Language Models and Generative AI

Huiwen Han
Lenovo IT
Lenovo
Beijing, China
hanhuiwen@gmail.com

*Abstract*: This paper introduces an innovative design for robotic operating platforms, underpinned by a transformative Internet of Things (IoT) architecture, seamlessly integrating cutting-edge technologies such as large language models (LLMs), generative AI, edge computing, and 5G networks. The proposed platform aims to elevate the intelligence and autonomy of IoT systems and robotics, enabling them to make real-time decisions and adapt dynamically to changing environments. Through a series of compelling case studies across industries including smart manufacturing, healthcare, and service sectors, this paper demonstrates the substantial potential of IoT-enabled robotics to optimize operational workflows, enhance productivity, and deliver innovative, scalable solutions. By emphasizing the roles of LLMs and generative AI, the research highlights how these technologies drive the evolution of intelligent robotics and IoT, shaping the future of industry-specific advancements. The findings not only showcase the transformative power of these technologies but also offer a forward-looking perspective on their broader societal and industrial implications, positioning them as catalysts for next-generation automation and technological convergence.

## I. Introduction

The Internet of Things (IoT) is reshaping industries by enabling the connection of devices, sensors, and systems into intelligent networks for real-time data collection, analysis, and decision-making. With projections indicating that the number of IoT devices will exceed 41 billion by 2025, the potential for technological advancements across various sectors is substantial. Robotics, as a key IoT application, is seeing rapid adoption in industries such as smart manufacturing[4][5], healthcare, and services. Traditionally, robots operated based on pre-programmed tasks, but the integration of IoT now allows them to dynamically adapt to environmental changes by analyzing real-time data and autonomously adjusting their operations.

In addition to environmental adaptability, recent advancements in large language models (LLMs) and generative AI have further enhanced robot intelligence. LLMs, like OpenAI's GPT and Google's BERT, allow robots to comprehend and generate human-like responses, fostering more intuitive interactions. This integration of generative AI with IoT provides robots with the capability to not only analyze data but also simulate and synthesize new data for decision support. For example, in smart manufacturing[4][5], generative AI can simulate production scenarios and create optimized models for predictive maintenance, helping to avoid equipment failure and improve resource allocation. Similarly, in healthcare, AI-generated simulations of patient data allow robots to predict patient conditions and autonomously adjust care plans, improving personalized treatment. In service industries, robots can synthesize customer preferences and behaviors to predict future demands and adapt their service delivery accordingly.

However, despite these advancements, several challenges remain. Issues such as data privacy, device interoperability, and standardization require attention to ensure seamless integration of these technologies. Moreover, maximizing the potential of generative AI in enhancing robot autonomy and adaptability is an ongoing research focus. This paper explores the synergy between IoT architecture and robotic operating platforms, examining the application of LLMs and generative AI in dynamic decision-making, environmental adaptation, and data simulation. It also discusses the future implications of these technologies across industries and the challenges that need to be addressed for their widespread implementation.

## II. Innovations in in IoT Architecture

The IoT architecture is foundational for achieving intelligent connectivity between devices, consisting of several layers, including the perception layer, network layer, and application layer, as depicted in Figure 1. Each layer plays a crucial role in the overall functionality of IoT systems, and the integration of advanced LLM technologies has significantly enhanced the intelligence of IoT systems[1].

## A. Components of IoT

### 1) Perception Layer

The perception layer is the foundation of the IoT architecture; it collects environmental data in real time through various sensors and edge devices, including temperature, humidity, lighting, motion, etc. To improve data processing at the perception layer, multimodal large model technology has gained attention in recent years. This technology can simultaneously process many types of data, including audio, video, and images, providing more comprehensive information understanding for IoT systems.

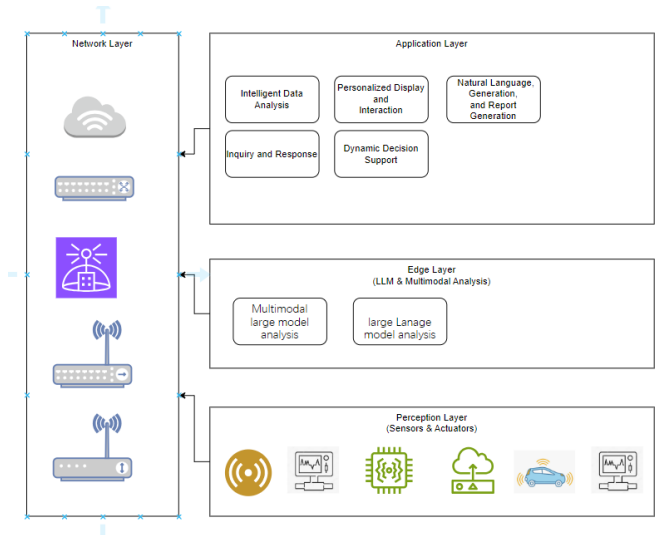

Figure 1. Basic IoT Architecture

For example, in a smart security system, the perception layer can capture scenes in real time through video surveillance cameras and collect environmental sounds using microphones. A multimodal large model can analyze visual information such as faces and movements in the video while combining audio signals to identify specific sounds, such as the sound of breaking glass, to quickly assess potential security risks and trigger alerts or other responses.

### 2) Network Layer

The network layer handles data transmission between devices, utilizing 5G and edge computing to ensure efficient and low-latency data transmission services. The high bandwidth and low latency of 5G networks enable simultaneous connections and data transmission for numerous devices, allowing IoT devices to respond and process data in real time.

By combining with large model technology, the network layer can use distributed computing platforms to process and analyze large-scale sensor data in real time. LLMs can provide semantic understanding and data conversion at this layer, thereby optimizing data transmission and processing flows. Through intelligent processing, the network layer can not only manage data flow more effectively but also generate real-time reports and decision-making recommendations after processing.

### 3) Application Layer

The application layer is the top layer of the IoT architecture, integrating various applications and services that use data collected from sensors for analysis and decision-making. The development of generative AI has significantly enhanced the intelligent decision-making and data analysis capabilities of the application layer. Combining LLM technology enables the application layer to perform more complex and intelligent functions.

## B. The Diversity of IoT Data

The Internet of Things (IoT), as a pivotal paradigm bridging the physical and digital realms, has profoundly reshaped various aspects of human production and daily life. Within this ecosystem, smart devices leverage sensing, communication, and computational technologies to generate a vast array of data types. The inherent complexity and extensive scope of IoT data have positioned it as a critical area of interest for both academic research and practical applications.

From a typological perspective, IoT data can be categorized into structured, semi-structured, and unstructured formats. Structured data, commonly sourced from sensor devices, encompasses quantitative metrics such as temperature, humidity, and pressure. Semi-structured data often appears in formats such as log files or XML-based information. Notably, the prevalence of unstructured data is increasing, with examples including video streams from surveillance cameras, audio recordings from microphones, and textual data generated through user interactions.

The diversity of IoT data is further exemplified by its wide-ranging sources, which span across domains such as smart homes[2][3], industrial automation, smart cities, healthcare, and precision agriculture. Each domain presents unique contextual attributes and data characteristics. For instance, traffic management systems in smart cities produce data fundamentally distinct from machinery diagnostic data in industrial automation environments.

In addition to its typological and domain-specific variations, IoT data exhibits multidimensional attributes. These include temporal characteristics such as real-time responsiveness, spatial dimensions relating to geographic distribution, and content heterogeneity reflecting varied formats and structures. Real-time data, such as traffic signal states, necessitates immediate processing, whereas geographically distributed data, such as environmental metrics from weather monitoring stations, requires spatially contextualized analyses, as depicted in Table 1.

In conclusion, the rapid proliferation of IoT technologies has engendered an unprecedented diversity in data forms and characteristics. This diversity not only fuels innovation in data-driven technologies but also presents significant analytical and operational challenges, thereby providing fertile ground for further research and development in the field.

## C. Applications of Generative AI in IoT-related Domains For Data Processing

Table 1.Types of data generated in different IoT-related application domains.

| Data | IoT-related Application Domains | | | | | |
|---|---|---|---|---|---|---|
| | Mobile Networks | Autonomous Vehicles | Metaverse | Robotics | Health Care | Cybersecurity |
| Vedio | ✓ | ✓ | ✓ | ✓ | ✓ | ✓ |
| Image | ✓ | ✓ | ✓ | ✓ | ✓ | ✓ |
| Audio | ✓ | ✓ | ✓ | ✓ | ✓ | ✓ |
| Text | ✓ | ✓ | ✓ | ✓ | ✓ | ✓ |
| Location | | | | | | |
| Device Status | | | | | | |
| Environment | | | | | | |
| Event- | ✓ | ✓ | ✓ | ✓ | ✓ | ✓ |
| Sensor | | | | | | |
| Usage | | | | | | |
| Vehicular Traffic | | | | | | |
| Network Traffic | | | | | | |

Applications of large models at the application layer include components, as depicted in Figure 2 :

### 1) Intelligent Data Analysis

Large models can perform in-depth analysis of massive datasets, identifying underlying patterns and trends. In IoT applications, this capability enables systems to detect anomalies in real time, predict future equipment demands, and optimize resource allocation. For instance, in the field of smart manufacturing, the application layer can use large models to analyze real-time data from production lines, enabling prediction of equipment failures and proactive maintenance to reduce downtime and maintenance costs.

### 2) Inquiry and Response

Large models can also provide natural language query functionality, allowing users to obtain information or perform actions using natural language. This feature, achieved through natural language processing (NLP), enables users to ask questions about device status, environmental parameters, or historical data, with the system parsing these queries and returning relevant information. For example, a user might ask, "What is the current temperature in my home?" The system, using a large model, would understand the question and immediately provide an accurate answer.

### 3) Dynamic Decision Support

With the integration of generative AI, the application layer can generate decision recommendations based on real-time data. These recommendations are not only based on historical data analysis but also adapt to real-time environmental changes. For example, in smart city management, the application layer can continuously analyze traffic flow data and generate optimal traffic signal control strategies to reduce congestion and improve urban efficiency.

### 4) Personalized Display and Interaction

Large-model technology can also enhance the user interface of IoT applications, making them more personalized and intelligent. Based on a user's historical behaviors and preferences, the system can automatically generate the information display that the user needs. Users can view various real-time data through an interactive dashboard and adjust the data display according to their specific requirements.

### 5) Natural Language, Generation, and Report Generation

In the IoT application layer, generative AI can be used to automatically generate reports and analytical results to help user understanding and decision-making. By analyzing the collected data, the system can automatically draft documents such as equipment performance reports and environmental monitoring results, reducing manual intervention and improving work efficiency.

### 6) Natural Language and Voice Control in IoT Devices

In the IoT application layer, natural language processing (NLP) and generative AI are used to enable seamless voice interaction with devices. These technologies allow users to issue commands and receive responses naturally, transforming user experience in smart environments.

### 7) Updated IoT Archietcture

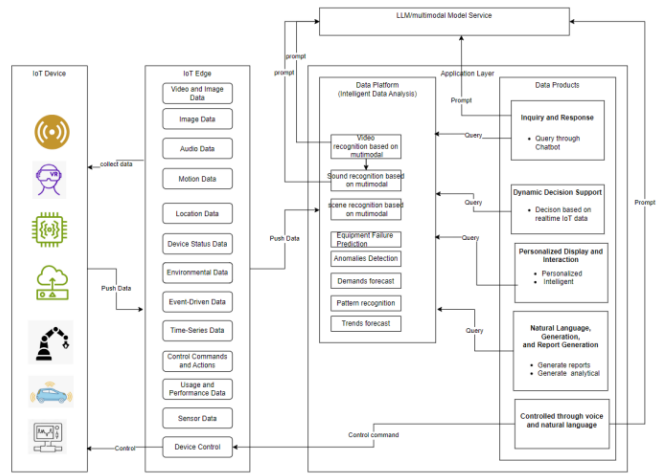

Figure 2. IoT Architecture with Gen AI for Data Processing.

## D. Applications of Generative AI in IoT-related Domains For Data synthesis and simulation

Generative AI has exhibited substantial advancements over traditional generative models, significantly enhancing the quality of generated content. For instance, contemporary models such as DALL-E 2[6] demonstrate the ability to produce highly detailed and visually compelling images. These outputs frequently achieve a level of realism and artistic sophistication comparable to, and occasionally indistinguishable from, photographs or artwork created by skilled professionals.

The integration of Generative AI into IoT-powered systems has positioned it as an indispensable tool across diverse domains, including mobile networks, autonomous vehicles, healthcare, and cybersecurity, as depicted in Figure 3.

*1) Mobile networks*

In **mobile networks**, Generative AI has revolutionized channel modeling and network traffic management. Conditional GANs have been employed for optical channel modeling, showcasing efficiency and precision. GAN-based frameworks have also been applied to air-to-ground channel modeling for UAV networks operating on millimeter-wave frequencies, facilitating distributed knowledge sharing and improving model generalization via federated learning. Additionally, GANs have transformed network traffic generation by simulating realistic transmission patterns, optimizing network scalability, and maintaining high performance with limited data. Innovations like WGAN-GP and Enhanced Capsule GAN have further advanced spectrum sensing by improving predictions of spectrum occupancy and channel estimation in IoT devices[8].

*2) Autonomous vehicles*

In the domain of **autonomous vehicles**, Generative AI enhances safety and operational efficiency through the creation of synthetic datasets, such as WEDGE, which improve vehicle perception under challenging conditions. These datasets are pivotal for real-time applications, including smart traffic control and vehicular Metaverse simulations, contributing to improved system reliability and responsiveness[8].

*3) Healthcare*

Generative AI has also driven transformative progress in **healthcare** through its integration with IoT devices. Applications range from monitoring patient vitals with wearables to classifying medical outcomes using GANs. For example, deep convolutional GANs have been employed in maternal and fetal health monitoring, while large language models (LLMs) like GPT-4 streamline clinical documentation, enhance patient interactions, and provide personalized care recommendations. Moreover, Generative AI improves electronic health records and facilitates humanoid doctors capable of interpreting IoT-collected data for diagnostic purposes[8].

*4) Cybersecurity*

In **cybersecurity**, Generative AI addresses the vulnerabilities of IoT systems by generating synthetic data that preserves essential statistical properties while mitigating privacy risks. Advanced models such as SecurityLLM and CySecBERT[7] leverage techniques like BERT and GPT-4 to detect threats, perform anomaly analysis, and enhance lightweight cryptography. These innovations significantly bolster the security of IoT devices, including those in vehicle networks[8].

*5) Digital Twin (DT) technologies*

Beyond these areas, Generative AI plays a critical role in **Digital Twin (DT) technologies**, enabling the creation of scalable virtual replicas of mobile networks to simulate user behaviors and optimize operations without affecting physical systems. In **robotics**, it facilitates continuous learning and behavioral modeling through frameworks like whole-brain probabilistic generative models, which emulate brain-inspired learning processes[8].

In conclusion, Generative AI's integration with IoT is driving substantial advancements across various industries, offering groundbreaking solutions for modeling, prediction, security, and efficiency. Its capabilities continue to shape the future of technology, unlocking new possibilities and addressing complex challenges.

*6) Updated IoT Architecture*

In IoT systems, acquiring diverse and high-quality datasets is often constrained by data sparsity, privacy regulations, and the difficulty of capturing rare or sensitive events, such as cyberattacks. Generative AI models, including large language models (LLMs) and multimodal systems, provide an innovative solution to this challenge. By leveraging raw data collected from IoT devices, these models can generate synthetic or simulated datasets that reflect real-world characteristics while overcoming limitations in data availability.

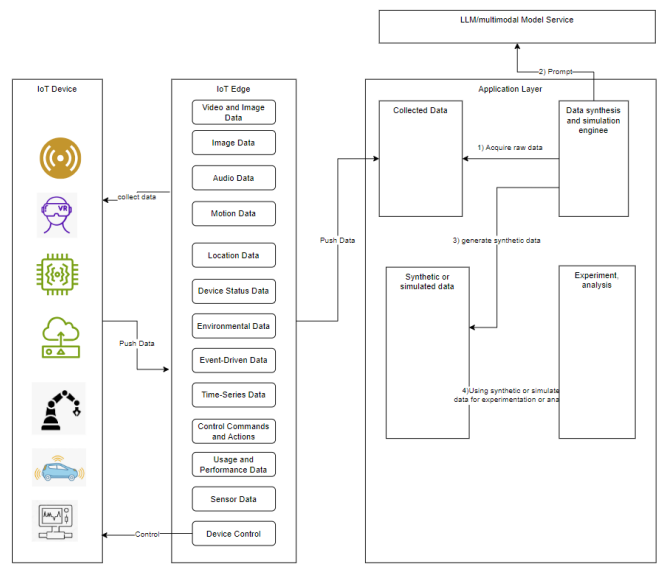

Figure 3. IoT Architecture with Gen AI for Data Synthesis and Simulation.

The process begins with the acquisition of raw IoT data, such as sensor readings, network logs, or telemetry data. This raw data serves as input to generative models, which are trained or fine-tuned on domain-specific scenarios to ensure the relevance and accuracy of the output. For example, in cybersecurity applications, synthetic datasets can simulate network attack patterns, enabling the development and testing of intrusion detection systems. Similarly, in smart healthcare systems, generative models can create anonymized patient datasets to support predictive analytics without compromising privacy.

This approach provides significant advantages, including the ability to simulate rare events, enhance privacy compliance, and reduce the cost and time required for real-world data collection. The synthesized data not only facilitates experimentation and model training but also enables scenario-based simulations, empowering researchers and practitioners to explore complex IoT applications in a controlled and scalable manner.

## III. INNOVATIONS IN ROBOTIC OPERATING PLATFORMS

### A. Definition and Development of Robotic Operating Platforms

The modern robotic operating platforms are not just control systems; they are evolving into intelligent decision-making and generation systems. With the integration of large language models and generative AI, these platforms can generate real-time responses and strategies, transforming robots from passive executors into proactive problem-solvers, as depicted in Figure 4.

*1) Innovative Features*
- Intelligent Interaction: By leveraging the natural language processing capabilities of large models, robots can interact with users in a natural, conversational manner, significantly improving the user experience.
- Dynamic Task Generation: Generative AI enables robots to create new tasks and working strategies in real time based on environmental changes and user needs. This feature allows robots to adapt to dynamic and unpredictable conditions, making them more versatile and resilient.

### B. Innovative Applications of IoT in Robotic Operating Platforms

*1) Remote Monitoring and Intelligent Feedback*

With 5G networks and generative AI, users can remotely monitor a robot's operating status in real-time, while AI-generated insights provide intelligent feedback and optimization suggestions. For instance, in smart manufacturing, the system can analyze production data and recommend optimized workflows, improving both efficiency and quality.

*2) Data-Driven, Intelligent Decision-Making*

Through IoT architecture and the analytical power of large models, robots can obtain real-time data for smarter decision-making. For example, in healthcare, service robots can use generative AI to create adaptive service plans based on patient needs, providing customized care solutions.

*3) New Modes of Collaborative Work*

IoT enables new collaboration modes for multi-robot systems. Using the reasoning capabilities of large models, robots can dynamically adjust work strategies based on shared information, achieving efficient collaboration, and improving overall performance in scenarios such as industrial automation and autonomous logistics.

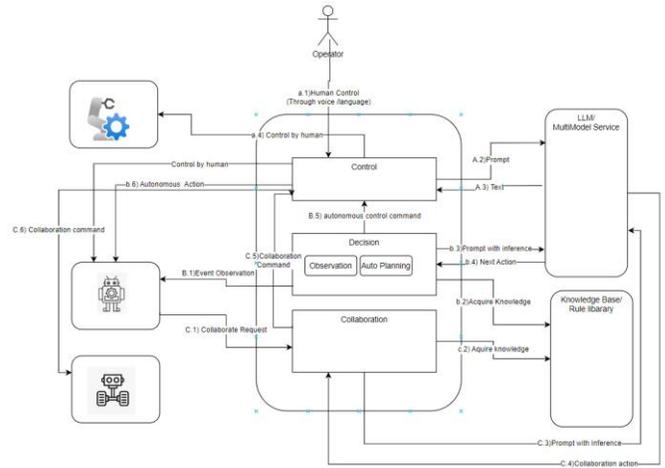

Figure 4. Robot Operation platform

## IV. CHALLENGES AND FUTURE TRENDS

While the combination of IoT and robotics presents broad potential, several challenges remain:

### A. Data Security and Privacy:

Enhanced measures for securing IoT devices and robot data are needed to protect sensitive information and ensure the safety of generative AI interactions.

### B. Standardization and Interoperability:

Establishing unified technical standards can facilitate interconnectivity between different devices and platforms, promoting broader adoption and smoother integration.

### C. Deep Integration and New Applications:

Future research should focus on the deep integration of IoT and generative AI, exploring applications in complex environments, and dynamic task generation to push the boundaries of robot autonomy and adaptability.

## V. CONCLUSIONS

The integration of IoT architecture with robotics establishes a transformative framework for advancing intelligent automation. By incorporating large language models (LLMs) and generative AI, robotic operating platforms can achieve unprecedented levels of contextual understanding, autonomy, and adaptability. This synergy drives real-time decision-making, predictive analytics, and user-focused interactions, unlocking novel opportunities across industries such as manufacturing, healthcare, and urban management.

However, the expansion of these technologies necessitates a parallel focus on security, privacy compliance, and governance. The increasing flow of sensitive data through IoT-enabled robotic systems demands robust cybersecurity protocols, strict adherence to data privacy regulations such as GDPR and CCPA, and comprehensive frameworks for ethical AI use. Furthermore, the integration must align with environmental, social, and governance (ESG) goals, ensuring sustainable development and equitable access to technological advancements.

Future research and innovation should prioritize secure and privacy-preserving architectures, transparent governance models, and systems designed for inclusivity and environmental responsibility. By embracing these principles, the convergence of IoT, robotics, and AI can catalyze scalable, ethically sound, and sustainable solutions that address global challenges while driving economic and social progress.

## REFERENCES

[1] Mahdi, Mustafa Jamal, Abbas Fadhil Aljuboori, and Mudhafar Hussein Ali. "Smart Stadium Using Cloud Computing and Internet of Things (IoT): Existing and New Models." *International Journal of Computer Applications Technology and Research* 10, no. 05 (2021): 111– https://doi.org/10.7753/IJCATR1005.1002.

[2] Y. M. Leong, E. H. Lim and L. K. Lim, "A Review of Potential AI-Based Automation for IoT-Enabled Smart Homes," 2023 IEEE 13th International Conference on System Engineering and Technology (ICSET), Shah Alam, Malaysia, 2023, pp. 1-6, doi: 10.1109/ICSET59111.2023.10295156.

[3] A. R. Dasgupta, P. Kumari, H. Sahu, S. Amarnath and S. Nanda, "Smart-Home Automation using AI Assistant and IoT," 2021 4th International Conference on Recent Trends in Computer Science and Technology (ICRTCST), Jamshedpur, India, 2022, pp. 329-333, doi: 10.1109/ICRTCST54752.2022.9781919.

[4] Z. -J. Zhou, D. -D. He, Z. -Y. Cao, J. -H. Kim, S. -Y. Kim and B. -L. Xu, "AI-Driven Predictive Diagnostic Maintenance for Lifespan and Efficiency of Massage Chairs," 2024 7th International Conference on Software and System Engineering (ICoSSE), Paris, France, 2024, pp. 61-65, doi: 10.1109/ICoSSE62619.2024.00018.

[5] M. Jide-Jegede and T. Omotesho, "Harnessing Generative AI for Manufacturing Innovation: Applications and Opportunities," 2024 International Conference on Artificial Intelligence in Information and Communication (ICAIIC), Osaka, Japan, 2024, pp. 568-572, doi: 10.1109/ICAIIC60209.2024.10463265.

[6] Ramesh, Aditya, Prafulla Dhariwal, Alex Nichol, Casey Chu, and Mark Chen. "Hierarchical Text-Conditional Image Generation with CLIP Latents." arXiv, April 12, 2022. https://arxiv.org/abs/2204.06125.

[7] Markus Bayer, Philipp Kuehn, Ramin Shanehsaz, und Christian Reuter.CySecBERT: A Domain-Adapted Language Model for the Cybersecurity Domain.arXiv preprint arXiv:2212.02974, 2022.

[8] Wang, Xin , Zhongwei Wan, Arvin Hekmati, Mingyu Zong, Samiul Alam, Mi Zhang, and Bhaskar Krishnamachari. "IoT in the Era of Generative AI: Vision and Challenges." arXiv, January 6, 2024. https://arxiv.org/html/2401.01923v2/.

.